\documentclass[twocolumn,american,prper,longbibliography]{revtex4-2}
\usepackage[LGR,T1]{fontenc}
\usepackage[utf8]{inputenc}
\usepackage{array}
\usepackage{amstext}
\usepackage{amssymb}
\usepackage{graphicx}
\usepackage{color}
\usepackage{soul}
\usepackage{float}
\usepackage{placeins}

\makeatletter

\newcommand{\lyxmathsym}[1]{\ifmmode\begingroup\def\b@ld{bold}
  \text{\ifx\math@version\b@ld\bfseries\fi#1}\endgroup\else#1\fi}

\providecommand{\tabularnewline}{\\}

\makeatother

\usepackage{babel}
\begin{document}
\title{Faculty Orientations Shape Adoption of AI in Research and Teaching}
\author{T. J. Atherton}
\email[Corresponding author: ]{timothy.atherton@tufts.edu}
\affiliation{Department of Physics \& Astronomy, Tufts University, 574 Boston Ave.,
Medford, MA 02155}
\author{I. Descamps}
\affiliation{Department of Physics \& Astronomy, Tufts University, 574 Boston Ave.,
Medford, MA 02155}
\author{T. R. Holmes}
\affiliation{Department of Physics \& Astronomy, University of Tennessee, Knoxville,
1408 Circle Dr, Knoxville, TN 37996 USA}
\author{C. L. Vizcarra}
\affiliation{Department of Chemistry, Barnard College, 3009 Broadway, New York,
NY 10027 USA}
\author{N. Sui}
\affiliation{Molecular and Structural Biochemistry, College of Agriculture and
Life Sciences, NC State University, Raleigh, NC 27695 USA}
\author{M. Webel}
\affiliation{Department of Physics \& Astronomy, Tufts University, 574 Boston Ave.,
Medford, MA 02155}
\author{J. J. Foley IV}
\affiliation{Department of Chemistry, University of North Carolina Charlotte, Charlotte,
North Carolina 28223, USA}

\begin{abstract}
Despite the widespread availability of large language models (LLMs) in higher education, instructors vary substantially in their adoption and use of these tools, and the reasons for this variation remain poorly understood. A mixed-methods survey of 90 STEM faculty in the Research Corporation for Science Advancement (RCSA) Cottrell community examined relationships between AI use, attitudes, institutional context, and instructional practice. Exploratory factor analysis identified a coherent construct, \textit{AI pedagogical orientation}, that strongly predicted self-reported AI use across research, teaching, and other professional activities. Qualitative analysis indicated that this construct reflected differing views about the role AI should play in disciplinary thinking, learning, and expertise development, rather than simply positive or negative attitudes toward AI. Institutional initiatives, demographic variables, and information sources showed comparatively weak associations with AI use. The results suggest that existing technology-adoption models may not fully explain adoption in contexts where technologies interact directly with disciplinary reasoning and knowledge production.
\end{abstract}

\maketitle

\section{Introduction}

The emergence of Large Language Models (LLMs) as a consumer-facing
AI product in 2022 has posed remarkable possibilities and challenges
for higher education. Providing a simulacrum of naturalistic dialog,
such models offer numerous use-cases including real-time interactive
search, topic exploration and explanation, and summarization. Agentic
models, which are able to execute tasks on behalf of the user, interact
with each other, and store information between sessions, can
facilitate delegation of complete tasks by users. Nonetheless, accuracy
of information provided by LLMs remains challenging, and concerns
around security, accuracy, and appropriate use remain. 

While dialogic LLMs are widely available, and are being
used by students and faculty alike, we presently lack understanding
of how LLM use shapes learning or disciplinary development. Nonetheless,
an initial landscape of educational applications and associated education
research is already emerging. Physics and Chemistry Education Researchers have begun
to investigate the capabilities of models on traditional problem sets,
as automated graders, virtual TAs or assistants that offer immediate, personalized feedback. There is also considerable
interest in applying LLMs to education research itself, potentially
to scale qualitative methods to larger data sets. An editorial accompanying
the recent Physical Review PER-focused collection on AI~\citep{Kuchemann2026} calls
for an integrated program to examine \textit{``how these tools reshape
what we measure, how students engage, how teachers work, and what
counts as understanding''}; an editorial in J. Chem. Educ. proposed a conceptual framework, CATALyST (Context, Applications, Technology, Attitudes, Learning, Skills, Tasks), oriented around similar goals~\citep{yuriev2024dawn}. It nonetheless remains an open question
whether adoption of LLMs will parallel adoption of other technologies into scientific
research and teaching, or whether distinctive pathways will emerge.

In this study, we examine how a group of pedagogically motivated STEM
faculty are using LLMs in their work. Within the emerging literature
on LLMs in education, a relatively small body of work has focused
specifically on instructors. Survey-based studies~\citep{Wattanakasiwich2025}
have examined patterns of adoption, identifying factors such as motivation,
perceived usefulness, and technical barriers as predictors of use.
Complementary qualitative work has begun to explore how instructors
make sense of generative AI~\citep{Skogvoll_2026}, showing that instructors
often hold multiple, overlapping interpretations of AI: as a useful tool,
a source of knowledge, and a potential threat to learning. Work beyond
physics and chemistry has examined instructors’ awareness of and attitudes toward
generative AI, finding generally positive sentiment but only weak
relationships with variables such as teaching style~\citep{Ghimire2024}.
Additional qualitative studies~\citep{desilva2026} document both
active adoption and cautious use, alongside concerns about student
learning and assessment validity, and the need for institutional support
and pedagogical adaptation.

Across these studies, instructors consistently recognize both the
possible uses and limitations of AI~\cite{lawrie2023assessment}. While these prior works document
both variation in adoption and the coexistence of multiple interpretations,
they do not fully explain why instructors respond differently to similar
affordances and concerns. Some instructors who adopt AI appear aware
of its limitations, while those who do not adopt it may still recognize
its potential utility. This suggests that variation in use cannot
be explained solely by access, information, or perceived usefulness,
as emphasized in existing models of technology adoption.

To understand possible mechanisms more clearly, here we determine how a selected group of pedagogically
motivated faculty, awardees of the Research Corporation for Science
Advancement (RCSA) Cottrell scholarship, are using LLMs in their work.
In the language of prior models of technology adoption, this group
represents a targeted population likely to be actively experimenting with AI in research and teaching, and variation
within this group offers insight into how other STEM faculty may respond
as these technologies become more widely adopted. 

To better understand these dynamics, we examine how this population uses AI tools in research, teaching,
and other aspects of their professional work. Using a mixed-methods survey of RCSA Cottrell awardees,
we identify an emergent construct, \textit{AI pedagogical orientation}, that helps
explain variation in AI use and reflects differing views about the role AI should
play in teaching, learning, and disciplinary practice. While the survey instrument
was designed to capture AI use broadly, LLMs emerged as the most widely reported
class of tools across the cohort. Accordingly, unless explicitly stated otherwise,
references to AI use in this work primarily concern LLM-based systems. 

The remainder of the paper situates this study within prior work on AI and computation in STEM education and existing theories of technology adoption (Sec.~\ref{sec:background}), describes the survey methodology (Sec.~\ref{sec:methods}), and presents the quantitative, qualitative, and mixed-methods analyses underlying the construct of AI pedagogical orientation (Sec.~\ref{sec:results}). We then discuss implications for instruction and faculty support, limitations of the present study, and directions for future work (Sec.~\ref{sec:discussion}) before concluding with broader implications for AI adoption in STEM education (Sec.~\ref{sec:conclusion}).

\section{Background}
\label{sec:background}

Recent work in Physics and Chemistry Education Research, as well as STEM education more
broadly, has begun to examine the implications of large language models
for instruction. Studies have evaluated model performance on traditional
physics problems~\citep{Tschisgale2025} and conceptual tasks~\citep{Polverini2025}, laboratory reports~\citep{west2023labreports}, chemistry problems involving multi-modal (graphical and structural) inputs~\citep{alasadi2024multimodal},
as well as students' responses to AI-generated solutions and instructional
materials~\citep{Dahlkemper2023,Lademann2025,ruff2024writing}. Other work has explored
the use of AI as part of instructional environments, including supporting
students in laboratory settings~\citep{Kilde2025} and acting as collaborative
partners in learning activities~\citep{Tong2025}, in lesson planning~\citep{aydin2025lesson}, in creating, refining, and evaluating student learning assessments~\citep{fernandez2024chatgpt,martin2023ml}, and even for facilitating
collaboration~\citep{Jiang2026}. Across these contexts, AI systems
demonstrate both productive capabilities and important limitations,
with studies consistently highlighting the continued importance of
human support, such as instructor oversight, in shaping how these
tools are interpreted and used in instructional contexts. More broadly,
studies beyond physics and chemistry education have found that the impact of LLMs
on productivity and performance is highly variable, with gains depending
strongly on task type and user expertise~\citep{noy2023experimental,brynjolfsson2025generative}. 
Indeed, LLMs can potentially both deepen and inhibit student engagement~\citep{lo2024influence}. 

Given that our understanding of AI in science education is at an early stage,
it is instructive to compare these efforts with the much longer history
of integrating computation into physics education, reviewed in Ref.
~\citep{atherton2023resource} with some emergent efforts described
in Ref. ~\citep{IOPBook}. Despite computation being viewed as important
for students' career preparation~\citep{McNeil.2017,Graves.2019},
and even central to physics knowledge production~\citep{phillips2023physicality},
large-scale survey data show that its integration has been uneven
and challenging~\citep{Caballero.2018}. Crucial to accelerated progress
in the last decade has been increased attention to faculty perspectives~\citep{young2019identifying},
identification of recommended practices~\citep{behringer2017aapt},
and the development of supportive learning communities such as the
Partnership for Integration of Computation into Undergraduate Physics
(PICUP) in response~\citep{Caballero.2019}. Similar efforts to create
communities in support of computation in chemistry education have also been formed with practices shared in the literature, see for example~\citep{mercury, stokes2021jupyter, Vizcarra2024, gardner2025pchem}.

At the same time, computation has been understood not simply as a
tool, but as a transformation of disciplinary practice. Work in Physics Education Research has framed computation as a means of conducting inquiry~\citep{burke2017developing,phillips2023physicality}.
Computation has also been framed as a form of disciplinary literacy~\citep{Odden.2019},
emphasizing its role in shaping how students engage in physics thinking.
While computation adds transformative possibility to learning environments~\citep{papert2020mindstorms,laurillard2020significance},
it also adds significant complexity in both instruction and curriculum
design. The incorporation of AI may similarly reshape disciplinary
practices while introducing comparable complexities, and early work
has begun to explore how AI influences such practices~\citep{fredly2026undergraduatephysicsstudentsuse}.
These perspectives highlight the importance of considering how AI
may reshape knowledge production in science and suggest that instructors’
interpretations of this role may be central to how such tools are
taken up in practice. Recent physics papers actively using AI for
creation of new physics theories~\citep{Seong2025} and materials~\citep{Wu2025}
illustrate that this transition is already happening in research contexts. 

Because there is not yet an established framework describing how instructors
integrate AI into disciplinary teaching practice, theories of technology adoption
provide a conceptual roadmap for identifying possible drivers of adoption. 
At the same time, if AI reshapes disciplinary practice, adoption may involve more
than decisions about its utility alone. We therefore use technology-adoption theories
both as interpretive frameworks and as points of comparison for understanding faculty engagement with AI. In designing and interpreting our study, we draw upon two contrasting
theories of technology adoption: Rogers' Diffusion of Innovations
(DoI)~\citep{rogers2003diffusion} and the Unified Theory of Acceptance
and Use of Technology (UTAUT)~\citep{venkatesh2003user}. Both frameworks
identify factors that are expected to influence adoption, and inform
the design of survey items and the interpretation of results in this
study.

The DoI theory conceptualizes adoption as a process involving the
innovation itself, communication channels, time, and a social system. 
In higher education contexts, these social-system dynamics could include disciplinary norms,
departmental culture, peer communication networks, institutional incentives,
and access to technical or pedagogical support, all of which can shape
faculty decisions about whether and how to adopt emerging technologies.
Factors that influence the rate of adoption include the relative advantage
of the technology, compatibility, complexity, trialability, and observability.
Based on these factors, innovations are expected to diffuse through populations over time,
from innovators and early adopters to an early majority, late majority, and those Rogers termed "laggards."
While widely used, including to understand faculty adoption of PER-based pedagogy~\citep{dancy2010pedagogical,henderson2012innovation,dancy2024physics},
this categorization has been critiqued~\citep{greenhalgh2004diffusion}
as unduly normative, as individuals may have well-founded reasons
for not adopting a technology. Moreover, adoption may not proceed
smoothly through these stages, but instead encounter barriers at multiple
points.

In contrast, UTAUT focuses on individual-level determinants of technology
use, emphasizing users’ perceptions and intentions rather than broader
patterns of diffusion. In this framework, adoption is predicted by
the perceived benefit of the technology, ease of use, social influence,
and facilitating conditions, with later extensions incorporating factors
such as motivation and prior experience. UTAUT has been widely applied
in educational contexts~\citep{xue2024unified,noureddine2025assessing},
where perceived usefulness and ease of use are often found to be strong
predictors of intention to adopt. However, like DoI, UTAUT primarily
treats adoption as a function of external conditions and user perceptions~\citep{bagozzi2007legacy},
and does not explicitly account for how individuals interpret the
role of a technology within disciplinary practice.

Comparing these perspectives, existing models of technology adoption
emphasize external conditions, user perceptions, and social context,
while research on computation in physics education highlights the
importance of how instructors conceptualize computers as part of disciplinary
practice. Extending this perspective to AI integration suggests that
instructors must not only decide \textit{whether} to adopt a tool,
but also \textit{how} to use it in teaching and research contexts.
Such decisions require instructors to conceptualize the epistemic
value of the tool---that is, how it may support disciplinary thinking
and what role it may play in knowledge production. How these broader adoption
dynamics interact with instructors' epistemic interpretations of AI remains
underexplored, particularly in explaining variation in faculty AI use.

\section{Methods}
\label{sec:methods}

To investigate how instructors' interpretations of AI relate to their
adoption and use of these tools, we conducted a mixed-methods survey
of a targeted population of STEM faculty. The study population, the
Cottrell community, consists of scientists who have received a competitive
award from the Research Corporation for Science Advancement (RCSA).
The Cottrell award is selectively given to junior faculty who apply
in the third year of a tenure-track job; applications are judged on
both quality of scientific research and an innovative education plan.
Due to the dual selection criteria of education and research, the Cottrell community
is a valuable population to study since it includes faculty likely to engage early
and critically with emerging technologies for research and instruction. At the time the survey
was conducted, the study population consisted of 572 individuals.

Design and distribution of the survey took place in Fall 2025. The
survey was designed by a core team of five STEM faculty from the CottreLLM
collaboration based on a sequence of structured discussions about
generative AI usage and adoption in research and teaching. Based on
these conversations, and also consultation with education researchers
who already published on AI usage, the survey scope was focused on
usage patterns across faculty activities and on factors that might
plausibly relate to or explain usage. Due to the limited size of the
Cottrell community, it was decided to adopt a mixed-methods design
and incorporate multiple free-response items to understand participants'
perspectives more clearly. Design proceeded iteratively, and the final draft
survey was shared with other members of the CottreLLM collaboration for feedback on
question phrasing. Additional input was solicited from two education
and AI researchers outside the Cottrell community as a cross check.
Small modifications were made in response to feedback to improve interpretability
by respondents. 

The survey was structured to collect data on factors drawn from technology
adoption and instructional change literature that are theorized to
influence adoption; the complete survey instrument is included in
the Appendix. Following screening and consent questions, an initial
section of three use questions asked how and whether the respondent
was using AI tools in research, teaching and the rest of their job;
these questions included associated free response questions about
how the participant was using these tools. A final free response question
asked what shaped their decision or any obstacles encountered. The
second section included questions on sources of AI tools the participant
used; the presence of institutional and department initiatives around
AI and respondents' agreement with a number of statements measuring
attitudes to AI. Respondents were also asked about their use of various
teaching methods and information sources they used to learn about
AI. The final section included basic demographic questions including career
stage, discipline and institutional type. Detailed demographic questions
were omitted to reduce the risk of respondent identification within
the relatively small and selective Cottrell community. 

Survey items were informed by constructs from Diffusion of Innovations
theory, particularly perceived attributes of the innovation (e.g.,
relative advantage and compatibility) and social-system influences.
Items measuring beliefs about AI’s pedagogical value align with Rogers’
construct of relative advantage, while items assessing comfort and
concern approximate perceived compatibility and complexity. Questions
regarding institutional encouragement and departmental initiatives
capture elements of the social system influencing adoption.

An invitation to participate in the research was distributed to all
members of the Cottrell community via an email listserv, with a second
reminder approximately four weeks after the initial contact. The survey
received a total of 116 responses total corresponding to a response
rate of approximately $20\%$. Responses were retained for further
analysis if at least $50\%$ of questions were completed to balance
data quality with sample retention in a limited population; this yielded
a total of 90 retained responses corresponding to $16\%$ of the population.

\section{Results}
\label{sec:results}

\subsection{Descriptive analysis}
\begin{figure*}
\includegraphics{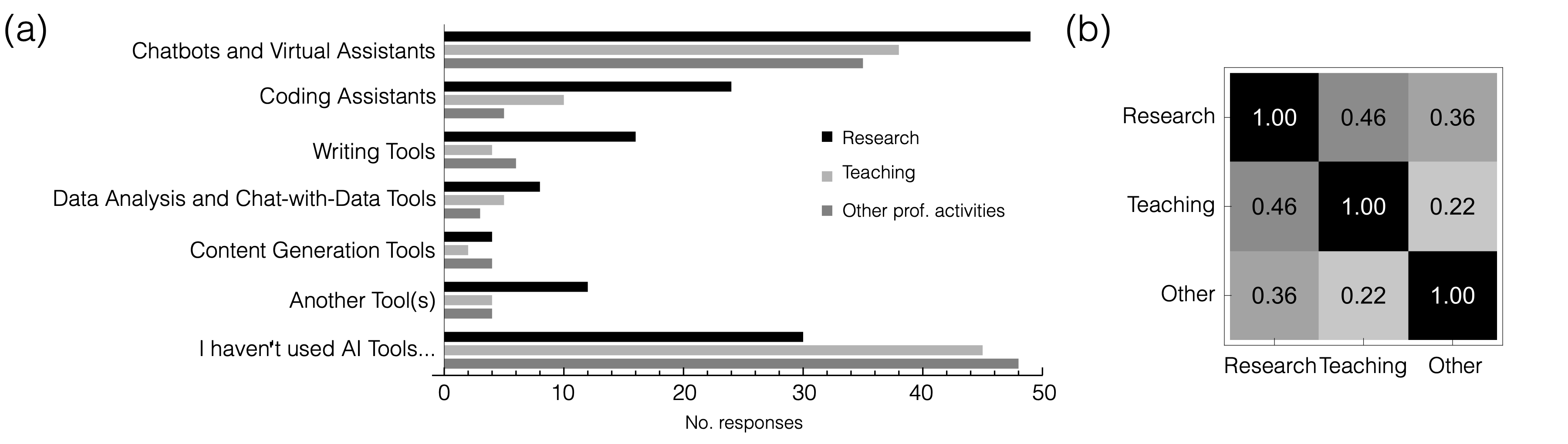}

\caption{\textbf{AI usage}. 
(a) Number of respondents reporting use of different categories of AI tools for research (black), teaching (light grey), and other professional tasks (dark grey).
(b) Correlation matrix showing pairwise correlations between binary indicators of AI use in research, teaching, and other domains.
}\label{fig:Usage}
\end{figure*}

We begin by reporting features of the dataset.  Responses to AI use questions indicate significant variation in usage
within the population (Fig.~\ref{fig:Usage}). While as many as $70$ participants
$(78\%)$ report using AI in some form and context, this is primarily
in the form of chatbots and virtual assistants ($49$ participants).
The second most common use is coding assistants, but much more so
in research $(24$ participants) than teaching or other aspects of
their job ($10$ and $5$ participants respectively). Other use cases
were much more limited. Free response questions typically referred
to particular products, e.g. \texttt{AlphaFold}, an AI model for predicting protein structures~\citep{jumper2021highly}, that respondents were using.

Overall, respondents were more likely to use AI tools in research,
but a significant number are also using AI in teaching and or other
aspects of their work. Collapsing usage into three binary indicators
for research, teaching and other aspects respectively, and computing
a correlation matrix between the three, we see that these use cases
are only partially correlated (Figure~\ref{fig:Usage}), research-teaching more so
than research-other or teaching-other. 

The usage data indicates a population that has begun to engage with
AI, in varied and domain-specific ways, but does not yet use AI uniformly
across all domains. Notably, $20$ participants $(22\%)$ reported
not using AI at all. 

\begin{table}
\begin{tabular}{lccc}
\hline 
 & Yes & No & Not sure\tabularnewline
\hline 
\hline 
Institution: degree / certificates & 27 & 31 & 32\tabularnewline
Institution: strategic initiatives & 58 & 13 & 19\tabularnewline
Department: degree / certificates & 9 & 70 & 11\tabularnewline
Department: strategic initiatives & 7 & 75 & 8\tabularnewline
\hline 
\end{tabular}

\caption{\textbf{AI initiatives} Summary of responses to survey questions about AI initiatives, certificates, and/or degree programs at the institutional and departmental levels.  Respondents represented Chemistry, Physics, and Astronomy Departments.}\label{tab:Institutional-and-department}
\end{table}

Institutional context questions (Table~\ref{tab:Institutional-and-department})
show that while 58 respondents $(64\%)$ report institutional strategic
initiatives around AI, only $27$ respondents ($30\%$) report degree
programs or certificates. In contrast, reported AI initiatives at
the departmental level were markedly less common: 9 respondents ($10\%$)
reported departmental degree programs or certificates, and 7 respondents
($8\%$) reported departmental strategic initiatives. These patterns
suggest uneven institutional engagement: many institutions have begun
to engage with AI, but these efforts have not yet fully been realized
in terms of programs, nor have they yet reached to the STEM departments
represented in this study. 

\begin{figure}

\includegraphics[width=1\columnwidth]{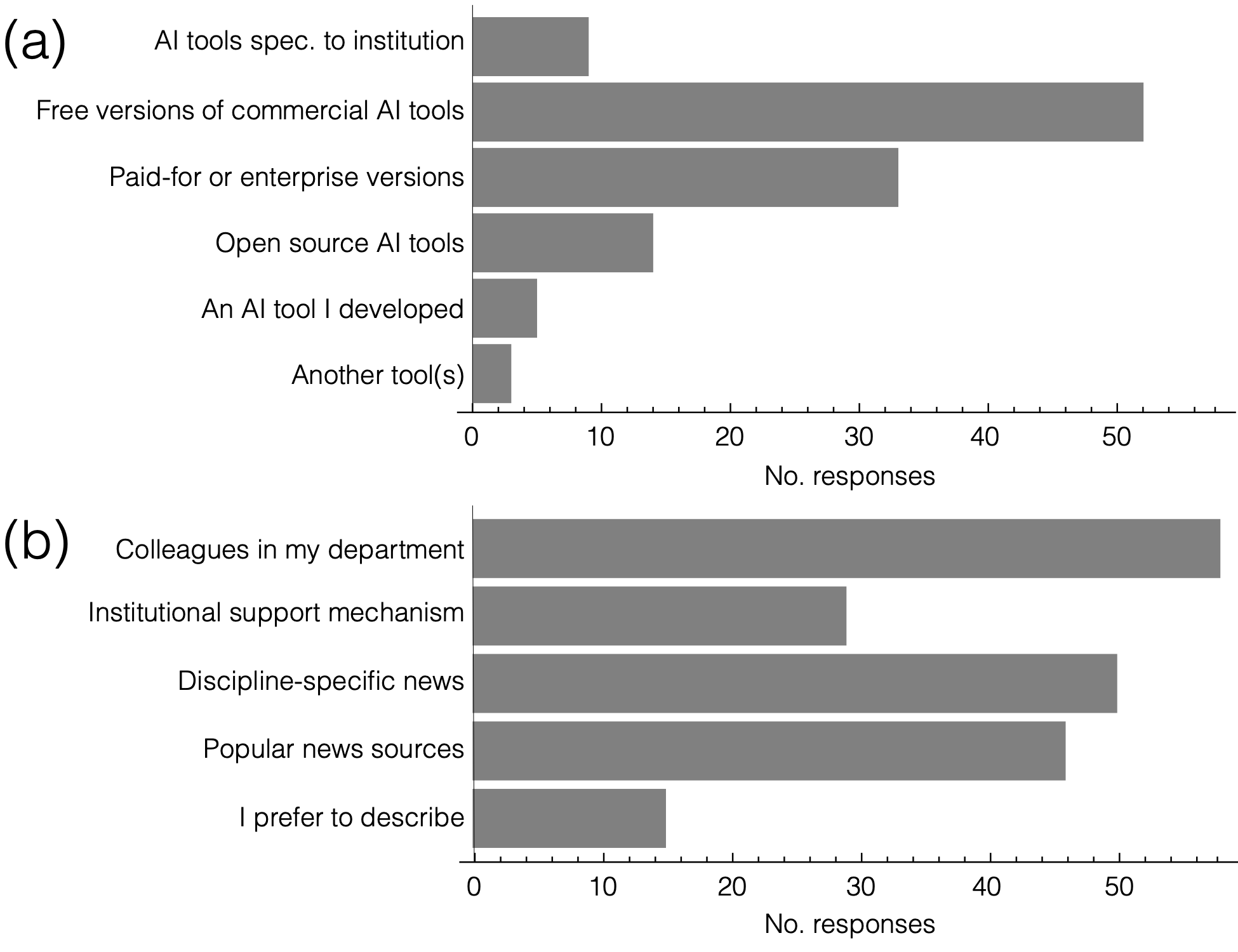}

\caption{\textbf{Sources and information about AI tools} (a) Number of respondents reporting use of different sources of AI tools.  (b) Number of respondents reporting reliance on specific data sources to learn about AI in higher education.}\label{fig:SourcesInformation}
\end{figure}

Figure~\ref{fig:SourcesInformation} displays results from the questions
about sources of AI tools. Some 52 respondents ($58\%$) reported
use of free versions of commercial tools; $33$ respondents ($36\%$)
use paid for or enterprise versions; use of open source or tools specifically
developed for an institution was selected less frequently. A small number of participants
report using AI tools they specifically developed, perhaps indicating
the presence of some individuals Rogers terms ``innovators'' in
the dataset. Also shown in Fig.~\ref{fig:SourcesInformation} are information
sources used by participants, which are most commonly department colleagues
($58$ participants or $64\%$) and news sources, whether discipline-specific
or the popular press. Institutional support mechanisms were used by
only 29 participants ($32\%$), suggesting 
that institutional initiatives have limited influence of faculty orientation in this cohort.

\begin{figure}
\includegraphics[width=1\columnwidth]{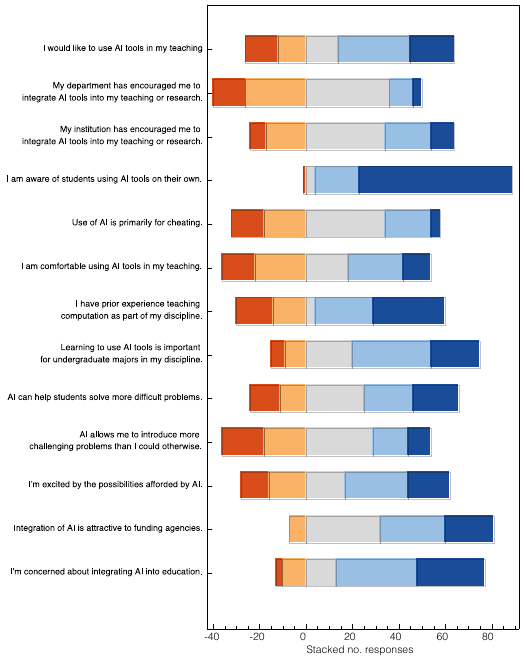}

\caption{\textbf{Distribution of responses to Likert-scale attitudinal statements about AI in teaching and research.} For each statement, colored segments indicate response categories: strongly disagree (red), disagree (orange), neutral (light grey), agree (light blue), and strongly agree (dark blue). 
The figure shows broadly positive attitudes toward AI’s pedagogical value coexisting with notable concerns about its integration.}\label{fig:Sentiment}
\end{figure}

Responses to attitudinal items displayed in Fig.~\ref{fig:Sentiment}
indicate generally positive but nuanced orientations toward AI, particularly
with regard to pedagogical value. Almost all participants report awareness
of student use, and most agree that learning to use AI is important
for physics and chemistry majors. Similarly, many respondents agreed that AI integration is attractive to funding agencies. Fewer report having
received encouragement from their department or institution, however,
and a significant minority feel uncomfortable using AI in teaching;
views on whether AI allows more challenging problems are similarly
mixed. While many respondents are excited by the possibilities afforded
by AI, nearly all are also concerned about integrating AI into education.
Overall, the sentiment data suggest a population that views AI as
pedagogically promising but is navigating uncertainty regarding institutional
support and appropriate integration.

\begin{figure}
\includegraphics[width=1\columnwidth]{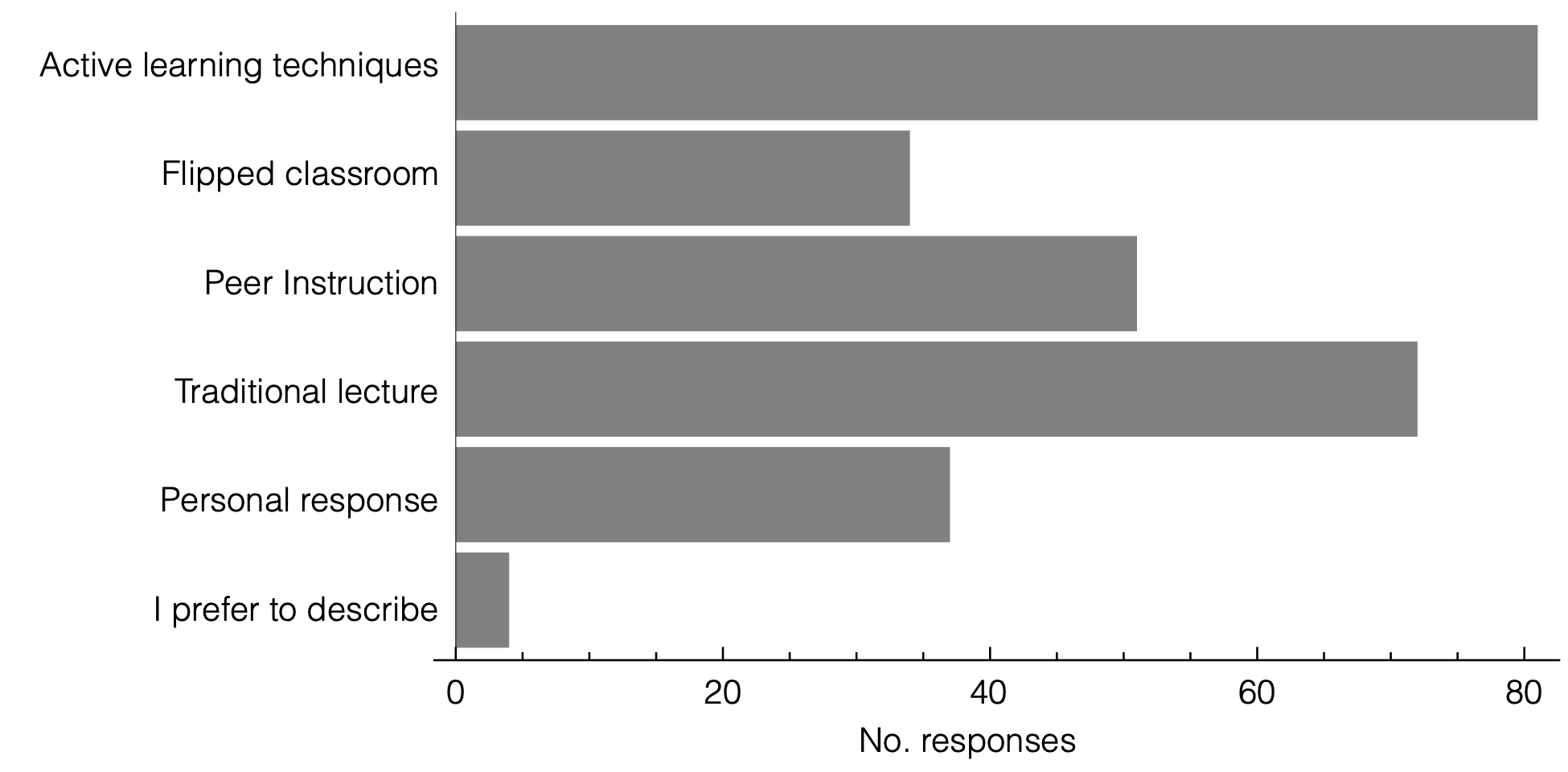}

\caption{\textbf{Instructional methods used.} Number of respondents reporting various methods in their teaching; these responses are independent of AI usage.}\label{fig:Teaching}
\end{figure}

Responses to questions on teaching (Fig.~\ref{fig:Teaching}) are
consistent with the highly engaged pedagogical profile of the Cottrell
community. More than $80$ respondents report using some form of active
learning, and a substantial fraction of participants use either flipped
classroom, personal response systems or peer instruction. A large
majority of ($72$ respondents) also report using traditional lectures,
suggesting that active and traditional techniques are being used in
conjunction or in different contexts. 

\begin{figure}
\includegraphics[width=\columnwidth]{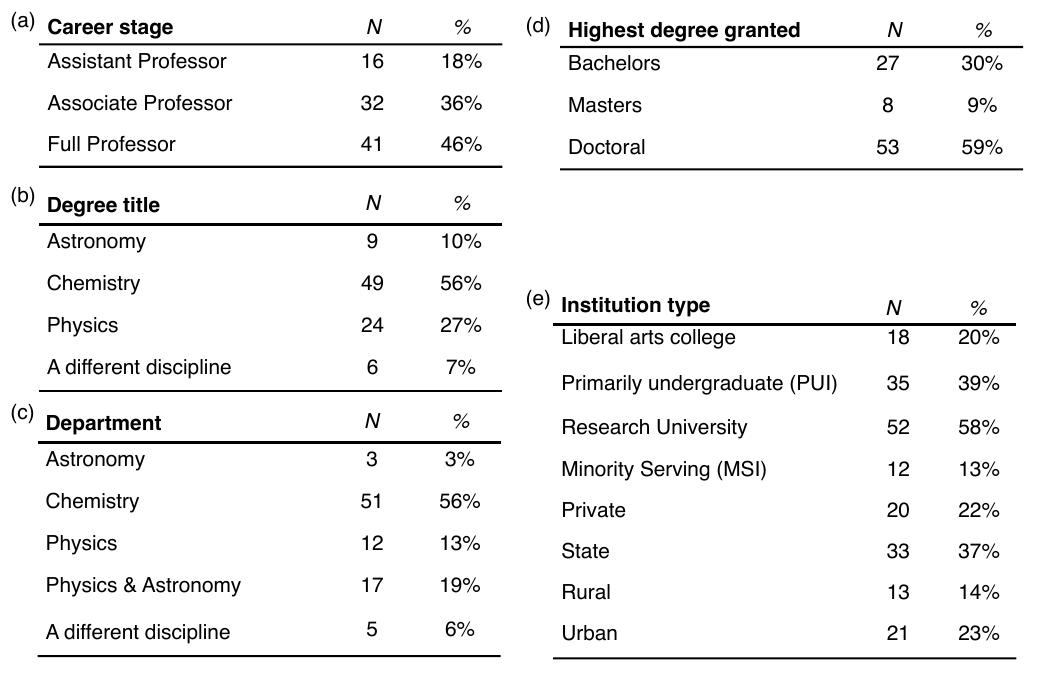}
\caption{\label{fig:demographics}\textbf{Respondent demographic and institutional information}. (a) Career stage of respondents. Note due to eligibility requirements of the Cottrell Scholar Award, all respondents are tenured or tenure-track faculty. (b) Respondent's degree. (c) Respondent's home department. (d) Highest degree granted by home department of respondents. (e) Respondent's institution type. Note that responses were non-exclusive.}
\end{figure}

Responses to demographic questions (Fig.~\ref{fig:demographics})
indicate a multidisciplinary population spanning a range of institution
types and career stages. Respondents include substantial fractions
at all career stages (Fig.~\ref{fig:demographics}(a)). Three disciplines,
Chemistry, Physics and Astronomy, are prominently represented both
in the original degree of the respondent (Fig.~\ref{fig:demographics}(b))
and their current department affiliation (Fig.~\ref{fig:demographics}(c)).
Responses to the ``prefer to describe'' option indicated some respondents
were originally trained in biological fields or had transitioned to
a biological or engineering discipline. 

As expected from the survey population, respondents are at a variety
of institution types and the sample includes a substantial fraction
from primarily undergraduate institutions (Figs.~\ref{fig:demographics}(d) and (e)); a smaller number report affiliation with
minority serving institutions. Taken together, these results are consistent
with the known composition of the Cottrell community. The presence
of respondents across multiple disciplines, career stages, and institution
types enables examination of whether AI engagement reflects individual
orientation alone or is patterned by disciplinary and institutional
context.

Descriptive analysis of the data reveals a population
that has begun to engage with AI in varied and context-dependent ways,
but use is uneven: there are differences between research and teaching
applications of AI, institutional encouragement is present but inconsistent,
and enthusiasm for AI’s pedagogical potential coexists with substantial
concern. These features suggest that AI adoption within this community
is not simply a matter of access or exposure, nor reducible to a single
positive-negative stance. Rather, the coexistence of engagement, ambivalence,
and structural variation raises the question of whether respondents'
beliefs and contextual experiences reflect coherent underlying orientations.
We therefore examine the latent structure of the attitudinal and contextual
variables using exploratory factor analysis, which we turn to in the
next subsection.

\subsection{Exploratory Factor analysis}

To examine the latent structure of the attitudinal and contextual
variables, we conducted exploratory factor analysis (EFA). Self-reported
AI use and demographic variables were excluded from the factor analysis,
as these were conceptually distinct from the remaining survey items
and were instead used for external validation of the resulting factors.

Survey data was prepared for analysis as follows: Survey items permitting
multiple non-exclusive responses (e.g., sources of AI tools, information
sources, teaching techniques) were recoded into binary indicator variables
for each response option. Institutional initiative questions were
recoded into two binary variables: an affirmation variable (Yes =
1; No/Not Sure = 0) and an awareness variable (Yes/No = 1; Not Sure
= 0), separating perceived presence from certainty. Binary variables
with extremely low (<5\%) or high (>95\%) prevalence were excluded
due to limited variance. Likert-type sentiment items (Strongly disagree
to Strongly agree) were treated as ordinal variables (1--5). After
pre-processing, the analytic dataset comprised $23$ binary and $13$
ordinal variables ($N=36$ total variables). Four missing responses
(<0.2\% of data points) were imputed using the median of the corresponding
variable. Excluded variables were retained for subsequent correlational
analyses.

Because the dataset included both binary and ordinal variables, we
computed a mixed correlation matrix using tetrachoric correlations
for binary--binary pairs, polychoric correlations for ordinal--ordinal
pairs, and polyserial correlations for mixed pairs~\citep{olsson1979maximum}. Exploratory factor
analysis~\citep{fabrigar2012exploratory} was conducted using minimum residual estimation with oblique
rotation. Analyses were conducted using the \texttt{psych} package in \texttt{R}~\citep{psych}.

Inspection of the eigenvalues of the correlation matrix and parallel
analysis with varying numbers of factors indicated the presence of
one dominant factor, with weaker evidence for additional factors.
Across models with $1-4$ factors included, a single factor consistently
emerged with stable loadings (the strength of association between each survey item and the factor) on a subset of variables. Despite the
robustness of the dominant factor, overall model fit remained poor
in initial solutions. We attribute this to the high variable-to-sample
ratio (36 variables; $n=90$) and the inclusion of heterogeneous binary
indicators that did not substantially co-vary with the factor.

To obtain a more interpretable solution, we conducted an iterative
refinement process, removing variables with low communalities and
minimal loadings on the dominant factor while monitoring stability
of the factor structure, reassessing model stability at each step.
This approach was guided by the goal of identifying a coherent latent
construct rather than maximizing overall model fit. Importantly, the
composition and interpretation of the dominant factor remained stable
throughout this refinement process; additional factors did not remain stable and hence were not analyzed further. The final EFA yielded a unidimensional
9-item scale with strong loadings ($0.45\lyxmathsym{–}0.94$) explaining
55\% of the variance. The factor showed strong internal consistency
(Cronbach’s $\alpha$ = 0.88, 95\% CI $[0.84,0.92]$; McDonald’s $\omega=0.89$)
and item-deletion analyses indicated no single item unduly inflated
reliability.

\begin{figure*}
\begin{centering}
\includegraphics{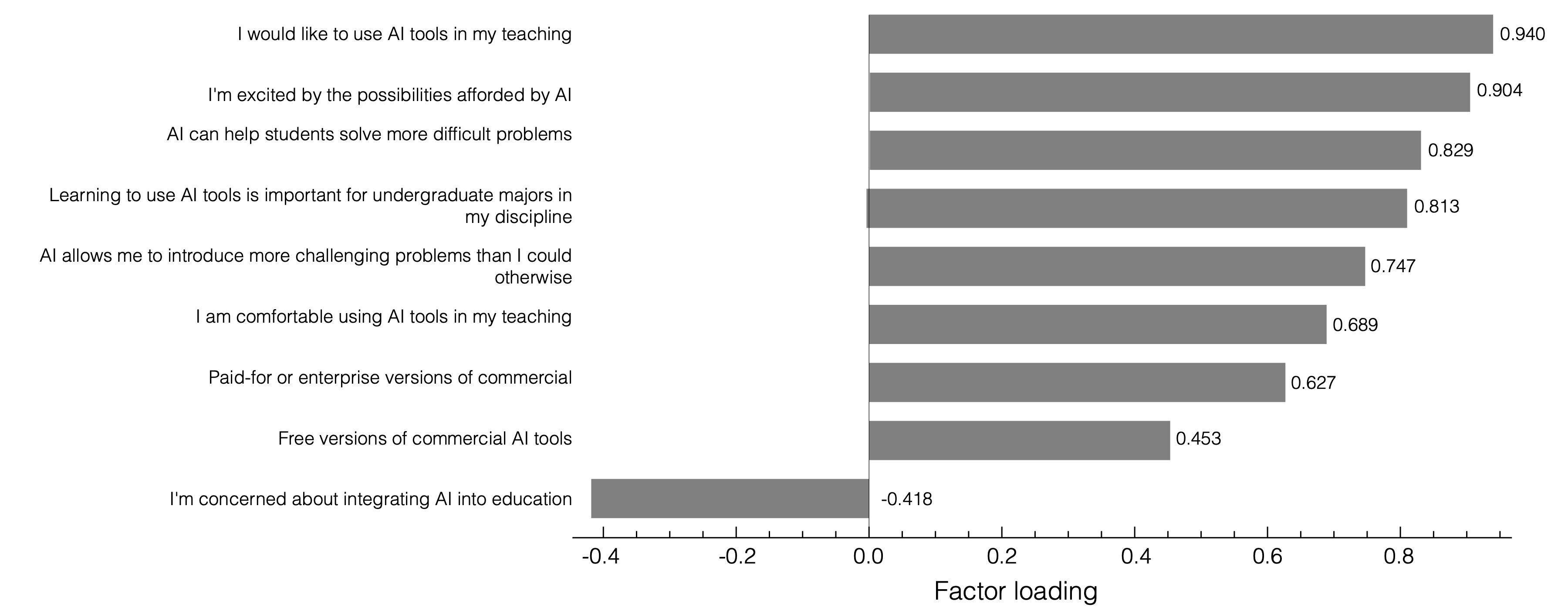}
\end{centering}
\caption{\textbf{Factor loadings for the ``AI pedagogical orientation'' factor} identified through exploratory factor analysis. Positive loadings (strength of association) indicate alignment with favorable attitudes toward AI use in teaching and learning, while the negative loading reflects concern about integration. Higher-magnitude loadings indicate stronger contributions to AI pedagogical orientation.}\label{tab:FactorLoadings}
\end{figure*}

Loadings (associations of variables with the identified factor) are displayed in Table~\ref{tab:FactorLoadings}. Items with the
strongest loadings ($>0.9$) reflected a desire to use AI tools in
teaching and excitement about their potential. Additional high-loading
items ($>0.8$) captured beliefs that AI is important for students
and enables engagement with more challenging problems. Moderate loadings
($<0.8$) reflected access to tools and comfort using AI. Concern
about integrating AI loaded negatively, consistent with the positive
sentiment of the remaining items. 

We interpret the identified factor as representing a participant's
\textit{AI pedagogical orientation}, capturing a coherent dimension
of beliefs about the role of AI in teaching and learning. The term \textit{orientation} is
intended to emphasize that the factor does not simply measure liking
or disliking AI, nor AI use alone. As explored in the qualitative and mixed methods sections below,
participants across the factor spectrum recognized both benefits and limitations of AI, and even respondents with strongly negative factor scores often reported some AI use. Instead, the factor appears to reflect differing stances toward how AI should be incorporated into teaching, learning, and disciplinary practice. The term \textit{pedagogical} reflects that many of the highest-loading variables concerned educational use and beliefs about student learning.

To examine whether the identified factor was associated with participants'
behavior, we conducted logistic regression analyses of self-reported
AI use with the factor score as the primary predictor. Results of
the regression for each of the three domains, research, teaching and
other professional tasks, are shown in Table~\ref{tab:LogisticRegression}.
AI pedagogical orientation was a statistically significant predictor
of AI use across all three domains (all $p<.001$). Log-odds estimates
and associated odds ratios show substantial effect sizes, with odd-ratios
ranging from approximately $2.8$ to $5.0$, with the strongest for
research and teaching. The consistency of these effects across domains
suggests that the construct reflects a general orientation toward AI
rather than domain-specific attitudes. Consistent with these results,
simple correlation between the usage variables and the identified
factor shows strong relationships between usage and the factor as
shown in Table \ref{tab:CorrelationUsageFactor}. 

\begin{table}
\begin{centering}
\begin{tabular}{lcccc}
\hline 
Outcome & $\beta$ & Odds Ratio & S. E. & $p$\tabularnewline
\hline 
Research & 1.61 & 5.00 & 0.34 & $2\times10^{-6}$\tabularnewline
Teaching & 1.45 & 4.26 & 0.31 & $4\times10^{-6}$\tabularnewline
Other & 1.03 & 2.80 & 0.27 & $1\times10^{-4}$\tabularnewline
\hline 
\end{tabular}
\par\end{centering}
\caption{\textbf{Logistic regression between binary usage variables and AI pedagogical
orientation factor}. $\beta$ coefficients are log-odds estimates;
odds ratios $e^{\beta}$ are also reported along with standard errors
and $p$-values from the regression. }\label{tab:LogisticRegression}
\end{table}

\begin{table}
\begin{centering}
\begin{tabular}{lc}
\hline 
Usage variable & Correlation with factor\tabularnewline
\hline 
Research & 0.594\tabularnewline
Teaching & 0.558\tabularnewline
Other & 0.442\tabularnewline
\hline 
\end{tabular}
\par\end{centering}
\caption{\textbf{Correlation between binary indicators of AI use} (research, teaching, and other professional tasks) and the ``AI pedagogical orientation'' factor. Positive correlations indicate that higher orientation is associated with greater likelihood of AI use across domains.
}\label{tab:CorrelationUsageFactor}
\end{table}

To further interpret the identified factor, we examined correlations
between the factor score and variables excluded from the EFA. Use
of institutionally developed tools, open-source tools, or self-developed
tools showed only weak associations ($\rho\le0.25$). Variables describing
institutional initiatives and awareness exhibited little consistent
relationship with the factor, with correlations generally near zero.
Similarly, teaching practices showed weak and inconsistent associations
($\left|\rho\right|\le0.19$), indicating that AI pedagogical orientation
is not strongly tied to general instructional approach. Notably, information
sources were also weakly related to the factor, with correlations
generally small and slightly negative ($\rho\in[-0.15,0.07]$). This
suggests that differences in AI pedagogical orientation are not explained
by differences in access to information or institutional support.

Examining sentiment variables excluded from the EFA, the factor showed
a moderate positive correlation with perceived departmental encouragement
to use AI ($\rho=0.47$), but other variables correlated weakly, including
institutional encouragement ($\rho=0.15$) and prior experience
teaching computation ($\rho=0.11$). Awareness of student AI use was
essentially uncorrelated with the factor ($\rho\approx0.00$). In
contrast, the belief that AI is primarily used for cheating showed
a substantial negative correlation with the factor ($\rho=-0.38$). 

We also examined correlations between the factor score and demographic
variables. Overall, relationships were weak, suggesting that AI pedagogical
orientation is not strongly associated with demographic characteristics.
For example, career stage showed no meaningful relationship with the
factor ($\rho=0.09,p=0.42$). Similarly, institutional characteristics
such as public/private status and minority-serving designation were
only weakly related to the factor. One exception was a moderate negative
correlation with rural institutions ($\rho\approx-0.34$) and liberal
arts colleges ($\rho\approx-0.25$), suggesting somewhat lower AI
pedagogical orientation in these contexts. However, given the small
sample size, this result should be interpreted cautiously. 

Taken together, these results indicate that AI pedagogical orientation is characterized primarily by positive beliefs about the pedagogical value of AI, enthusiasm regarding its educational possibilities, and comfort with its use in teaching. The factor is also associated with participants' epistemic interpretation of AI, particularly whether it is viewed as supporting or undermining learning, and availability of tools. In contrast, it appears largely independent of demographic factors, institutional context or messaging, teaching practices, information exposure, and awareness of student behavior.

\subsection{Qualitative analysis}

To better understand how these differences in AI pedagogical orientation
are expressed in instructors’ reasoning and experiences, we turn to
qualitative analysis of open-ended survey responses (Q2, Q4, Q6, Q7,
Q33). Qualitative analysis was performed using an iterative, emergent
coding procedure. The responses for each question were de-identified
and circulated to an initial team of five coders who independently
reviewed the data to identify emergent themes. The coders subsequently
met to synthesize their observations into an initial set of 25 tentative
codes (see Table~\ref{tab:ai_codebook} in appendix). Using this codebook, each coder
then coded each response to indicate which codes were present. Codes
were non-exclusive; a single response could receive multiple codes
or none.

Inter-rater reliability for the initial code set was assessed using
Krippendorff’s $\alpha$ for each code. Two codes, \textit{nomenclature}
and \textit{coding}, exceeded $\alpha>0.8$ demonstrating high reliability.
A further six, \textit{learning concerns}, \textit{visualization},
\textit{resource gathering}, \textit{ethical}, \textit{writing} and
\textit{reliability}, met the commonly accepted threshold of $\alpha>0.67$
indicating tentative reliability. Five additional codes approached
this threshold $\alpha\gtrapprox0.6$ and were identified as candidates
for potential refinement.

Rather than refining the exploratory code set, we observed that the
emergent codes could be grouped into four higher-order constructs
that align with theoretical perspectives on technology adoption and
learning. Below we list each code and the associated emergent codes: 
\begin{enumerate}
\item \textbf{Practice. }Respondent describes concrete uses, workflows,
or tasks in which they personally used (or directly supervised use
of) AI tools. \textit{Emergent codes: Coding, Visualization, Resource-gathering,
Writing, "Administrata", Generation, Refinement.}
\item \textbf{Epistemic.} Respondent focuses on AI as an epistemic object---its
reliability, evaluation, limitations, how to validate outputs, how
it affects thinking/learning, or how it supports inquiry. \textit{Emergent
codes: Ideation, Basic-info, Assessing-AI, Exploration, Validation.}
\item \textbf{Affective.} Respondent expresses emotions, motivation, value
judgments, comfort/discomfort, excitement, fear, ethical worry, or
personal stance toward AI. \textit{Emergent codes: Sentiment-positive,
Sentiment-negative, Empathy, Ethical, Learning-concerns.}
\item \textbf{Structural.} Respondent references institutional, departmental,
or disciplinary structures that shape AI use: policies, initiatives,
programs, support mechanisms, governance, access constraints, or organizational
norms. \textit{Emergent codes: AI-development, End-user, Obligation,
Accessibility, Nomenclature, Institutional.}
\end{enumerate}
These four categories also map directly onto patterns observed in
the quantitative analysis, particularly the distinction between epistemic
and structural influences on AI adoption.

\begin{table}
\begin{centering}
\begin{tabular}{lccc}
\hline 
Code & $\alpha$ & Coder A & Coder B\tabularnewline
\hline 
Practice & 0.83 & 151 & 135\tabularnewline
Epistemic & 0.67 & 76 & 76\tabularnewline
Affective & 0.46 & 30 & 67\tabularnewline
Structural & 0.45 & 11 & 29\tabularnewline
\hline 
\end{tabular}
\par\end{centering}
\caption{\textbf{Results of consolidated coding} showing Krippendorff’s $\alpha$ and
the number of examples found by each coder. Differences in counts
for Affective and Structural codes reflect variation in coding thresholds
rather than conceptual disagreement.}\label{tab:Coding2}
\end{table}

We therefore recoded the full dataset using this consolidated framework.
Two new coders, who had not participated in the initial emergent phase,
independently applied the four higher-order codes to all responses;
this procedure ensured analytic separation between exploratory theme
generation and confirmatory recoding. After an initial round of independent
coding, the coders met to compare interpretations and clarify code
definitions; they then had the opportunity to revise their coding.
This process was intended to improve consistency in code application
while preserving independent judgment.

Inter-rater reliability for the reduced code set was
assessed using Krippendorff’s $\alpha$ (Table \ref{tab:Coding2}).
The Practice code showed high reliability $\alpha=0.83$, and the
Epistemic code met the accepted threshold for tentative reliability
$\alpha\approx0.67$. The remaining two codes, Affective and Structural,
exhibited lower $\alpha$ values ($0.46$ and $0.45$ respectively).
Inspection of coder agreement indicated that this lower reliability
was primarily driven by differences in coding thresholds rather than
conceptual disagreement. Nearly all instances identified by the more
conservative coder were also identified by the second coder (e.g.,
28 of 30 Affective instances and 10 of 11 Structural instances), indicating
strong agreement on clear cases but divergence in more ambiguous instances.
We therefore interpret Affective and Structural codes as reliably
identifying explicit instances of these constructs, while acknowledging
greater variability in marginal cases. For subsequent joint analysis,
we adopt a conservative approach and treat codes as present only when
both coders independently identified them.

\subsection{Joint analysis}

To understand how AI pedagogical orientation is reflected in respondents'
reasoning, we examined relationships between the consolidated qualitative
codes and the factor score computed above. Results are displayed in
Table~\ref{tab:CorrelationMixed}. Correlations were computed using
two approaches: in the first column, codes were based on the consolidated
coding described above; in the second, correlations were computed
using the original emergent codes, grouped into the consolidated categories
using a majority-rule criterion. Both approaches yield highly consistent
results.

\begin{table}
\begin{centering}
\begin{tabular}{lcc}
\hline 
Code & Correlation & Original codes\tabularnewline
\hline 
Practice & 0.467 & 0.554\tabularnewline
Epistemic & 0.222 & 0.232\tabularnewline
Affective & 0.103 & -0.071\tabularnewline
Structural & 0.080 & 0.070\tabularnewline
\hline 
\end{tabular}
\par\end{centering}
\caption{\textbf{Correlation of consolidated qualitative codes with factor score.}
Correlations were computed with the recoded data and using the initially
coded data with emergent codes grouped into consolidated codes. }\label{tab:CorrelationMixed}
\end{table}


The strongest relationship was observed for Practice codes, which showed a substantial positive correlation with the factor score across both methods. Epistemic codes showed a weaker but consistently positive association, while Affective and Structural codes exhibited little systematic relationship with the factor. Although the strong relationship with Practice codes might be anticipated given that these codes primarily captured descriptions of concrete AI use, we emphasize that the association emerges through the integration of independently derived quantitative and qualitative analyses. In contrast, Epistemic codes encompassed both supportive and skeptical reflections on AI’s role in learning and reasoning, which may contribute to their weaker overall association with the factor. The results suggest that higher AI pedagogical orientation is associated most directly with reported engagement in AI-related practices, while epistemic reflection appears more heterogeneous and not uniformly aligned with positive orientation toward AI.

To further contextualize the factor, we examined quotes from respondents
with the ten most negative and most positive values of the factor
score. Responses from participants with the most negative factor scores
were characterized by limited and cautious use of AI, coupled with
strong epistemic concerns about its impact on learning. When AI was
used, it was typically described as peripheral or narrowly scoped.
For example, one respondent described their use of machine learning
as \textit{``a very minor and exploratory aspect of my research,''}
while another reported using ChatGPT only for \textit{``help configuring
the environment of one of my work computers.''} Even when AI proved
useful, it was framed cautiously: \textit{``AI wasn't a magic bullet
but it did eventually point me in the right direction.''}

The responses of those participants most oriented away from AI 
contained many examples of reflections on the role of effort and deliberate
engagement in learning. Several respondents articulated a clear model
of learning that emphasized depth over efficiency. As one participant
explained, \textit{“deep learning is inherently a slow process that
cannot be done quickly,”} requiring \textit{“sufficient time engaging
with cornerstone concepts.”} From this perspective, AI---and particularly
language models---was seen as promoting superficial engagement. One
respondent noted that students may \textit{“quickly find the specific
nugget of knowledge they need to check a box,”} but \textit{“have
not engaged the deeper neural networks that lead to authentic learning,''}
resulting in knowledge that is \textit{``superficial and in a few
weeks the slate is clean.''}

These concerns were often accompanied by explicit epistemic boundary-setting
around what counts as acceptable AI use. Multiple respondents distinguished
between forms of AI, expressing acceptance of machine learning for
data analysis while rejecting LLMs as problematic. For example, one
respondent emphasized that \textit{“scientists use machine learning
and neural networks\ldots{} {[}that{]} are objective statistical tools,”
}in contrast to LLMs, which were described as \textit{“computers pretending
to be people.” }Another noted that \textit{“not all AI is LLMs,” }and
that different forms of AI have \textit{“very different societal impacts.”}

Across responses, AI was frequently positioned not simply as a tool,
but as a potential threat to the development of disciplinary thinking.
Concerns centered on over-reliance, loss of critical evaluation, and
the erosion of deep learning practices. As one respondent cautioned,
\textit{“if you can get AN answer quickly, but not decipher whether
it is a GOOD answer, you run the risk of overconfident incorrectness.”}
Others raised broader concerns about the direction of educational
change, suggesting that AI initiatives risk \textit{“putting the cart
before the horse”} by prioritizing efficiency over understanding.

In contrast, responses from participants with the highest factor scores
were characterized by extensive and integrated use of AI across research,
teaching, and professional tasks. Rather than describing isolated
or occasional use, these respondents portrayed AI as embedded within
their daily workflows. For example, one respondent described using
AI for \textit{``data analysis, plotting, coding in various languages\ldots{}
first-draft of text passages, outline generation of articles, {[}and{]}
summary of references,”} while another reported using AI to \textit{“brainstorm
research ideas\ldots{} conduct literature reviews rapidly\ldots{} outline
skeleton versions of code, and occasionally debug small snippets of
code.”} Across responses, AI was presented as a general-purpose tool
supporting a wide range of disciplinary and professional activities.

A common theme in the high factor score responses was the role of
AI in increasing efficiency and reducing time spent on routine tasks.
Many respondents emphasized the productivity gains afforded by AI,
describing it as a tool that could \textit{“save time”} or dramatically
accelerate their work. In some cases, these gains were framed in dramatic
terms, with one respondent reporting that AI could \textit{"boost
your individual productivity by 5--10x," }while another described
the difference as\textit{ "night and day"} compared to working without
it. AI was frequently used to streamline writing, administrative work,
and technical tasks, with one respondent noting that tasks that \textit{“used
to take an hour\ldots{} now take{[}{]} 30 seconds.”} This perspective positions
AI not simply as a tool for solving problems, but as a means of redesigning
professional workflows.

Beyond efficiency, many respondents described AI as an interactive
cognitive partner that supports thinking and idea development. Several
reported engaging in \textit{"conversations"} with chatbots to refine
arguments, explore research directions, or structure written work.
AI was described as providing \textit{“immediate feedback on ideas
and methods,”} enabling rapid iteration and exploration. This perspective
extended into teaching, where instructors described incorporating
AI into assignments, generating instructional materials, and designing
activities in which students critically engage with AI-generated content.
For example, one respondent described having students \textit{“look
for inaccuracies in the LLMs\ldots{} {[}to{]} teach them how to use
LLMs in a constructive, and critical way.”} In this view, AI is
not only a productivity tool but also a resource for supporting thinking.

Importantly, these respondents were not uncritical and demonstrated
awareness of the limitations and potential risks associated with AI
use. Many noted that AI outputs \textit{“don’t always get it right”}
and must be used \textit{“with care,”} emphasizing the need for verification
and critical evaluation. Some explicitly acknowledged trade-offs between
efficiency and understanding, with one respondent noting \textit{“a
trade off between my increased productivity and my decrease{[}d{]}
ability to think by myself.”} However, rather than leading to rejection
or avoidance, these concerns were framed as issues to be managed.
As one respondent observed, \textit{“instead of focusing on whether
we should or shouldn't use AI, we should focus on how to use it so
that we do not lose our ability to think.”} This suggests that, for
high-orientation respondents, epistemic concerns coexist with---and
are incorporated into---ongoing use of AI.

The contrasting patterns between respondents with the most negative
and most positive factor scores suggests that the AI pedagogical orientation
factor reflects differences in how instructors interpret the role
of AI in thinking. While both groups recognize potential limitations
and risks, respondents with negative scores frame these concerns as
reasons to limit or avoid AI use, whereas those with positive scores
treat them as manageable constraints and integrate AI into their workflows.
This distinction is consistent with the quantitative results, in which
the factor strongly predicts AI use but shows little association with
institutional or informational variables.

\section{Discussion}
\label{sec:discussion}

\subsection{Proposed model of AI adoption}

This study identifies a coherent construct, \textit{AI pedagogical
orientation}, describing how instructors position AI within teaching, learning, and disciplinary practice. 
The construct strongly predicts respondents' use of AI across
research, teaching and other professional activities. The construct
reflects both sentiments around AI as well as ready access to AI tools.
Notably, a positive orientation does not simply reflect an uncritical
liking of AI, nor does a negative orientation represent a rejection:
rather, respondents across the spectrum recognized both benefits and
concerns about AI. 

Our findings differ in important ways from existing accounts of technology
adoption. Models such as diffusion of innovations and the UTAUT have
emphasized factors such as access to tools, social influence and perceived
utility. We do see some evidence for these factors: access variables
comprise part of the pedagogical orientation construct, for example.
Positively oriented respondents frequently discussed utility, but
negatively oriented respondents also saw value in the technology despite
their decision not to adopt it. Further, information sources and institutional
initiatives showed little association with AI usage, at least in this
population and at this stage in adoption of the technology. Institutional
and demographic variables similarly showed only weak, contextual variation.
Instead, variation in AI usage seems to be most strongly associated
with instructors' conceptualization of AI in disciplinary thinking. 

We therefore propose a model of AI adoption where epistemic interpretation
of AI shapes the relationship between sentiments of AI and its practical
use. In this model, faculty differ not in whether they see potential
benefits and risks around AI, but in how they interpret the role of
AI in disciplinary thinking, and knowledge production more broadly.
For those oriented away from AI, it is seen as potentially undermining
the careful engagement necessary to develop expertise, leading to
limited or cautious use. For those oriented towards AI, the ability
to potentially delegate thinking---especially for tasks seen as ancillary---in
support of higher level thinking and exploration is attractive, with
risks treated as constraints to be mitigated through critical use.
AI pedagogical orientation reflects not simply attitudes toward a technology, 
but differing views about the role AI should play in disciplinary thinking, learning, and expertise development.
This proposed model is consistent with our results. AI pedagogical
orientation strongly predicts AI usage across research, teaching and other professional activities, while other
variables show weak associations. Future work should pay attention to 
whether the external factors currently not driving adoption become relevant
at later stages and as technologies, information and institutions
mature in their response.

\subsection{Implications for instruction}

Findings from this study have a number of implications for how AI
is introduced and supported in STEM higher education. Most importantly,
they suggest that efforts to support AI adoption must help faculty
conceptualize and articulate how AI might support disciplinary thinking
in their context, as well as constraints necessary so that important
developmental stages are not short-circuited. Respondents differed
in how they interpreted benefits and limitations of AI, and hence
efforts that focus solely on providing access and technical training
may be insufficient if they do not also engage in participants' beliefs
about thinking and learning. 

Rather than positioning AI as a tool to be adopted or avoided, instructors
may benefit from thinking of AI in the context of a broad set of practices
to support reasoning and problem solving, particularly those that
could make its potential role in thinking explicit. Such support might
include developing strategies for evaluating AI outputs, identifying appropriate contexts for delegation of thinking, and designing tasks that require validation and revision of AI-generated work. This aligns well with how respondents positively
oriented towards AI are integrating them thoughtfully into their disciplinary
practice, combining efficiency with evaluation and reflection. 
This also aligns with recent PER work on student agency, which frames interactions with computational and disciplinary tools as opportunities for scientific decision making, modeling, and epistemic agency~\citep{holmes2020developing,smith2020expectations,phillips2023physicality}. 

Curricula and instructional design should aim to provide students
with opportunities to critically engage with AI and AI generated-content
in ways that align with emerging disciplinary use. Concerns around
AI often centered on the potential for superficial understanding,
highlighting the need for tasks that require validation, interpretation,
robustness checking and revision of work, including AI-generated output,
as well as comparison between AI and human-generated work. With
these, opportunities for AI to support ideation, exploration and provide
rapid feedback---as some respondents are already doing in their professional
scientific work---can be incorporated without displacing disciplinary
thinking. Indeed, AI-supported instruction may create additional opportunities to
make student thinking visible, in line with longstanding PER goals~\citep{hammer2000student}.

At the institutional level, the weak association between AI use and
institutional initiatives observed here suggests that policies and
support structures in isolation may have limited impact on adoption.
As these institutional structures become more established and embedded
in schools and departments, they may make most impact if, rather than
simply providing information, resources or establishing guidelines,
they instead incorporate opportunities for discussion and reflection
on the issues at stake, as well as low-stakes opportunities for situated
disciplinary use of AI . We note that in the present sample, faculty
were likely to rely on other faculty in their department as information
sources, so models that lean into distributed support networks leveraging
discipline-specific knowledge rather than centralized efforts may
prove more successful. 

\subsection{Limitations and future work}

We note a number of limitations to the present study. While the sample
is a significant fraction of the target population, RCSA Cottrell
awardees, the overall sample size is modest. Nor is the population
(by construction) intended to be representative of the broader group
of STEM faculty: Use of contemporary pedagogical practices is higher
in this sample than in the general STEM faculty population, for example,
where traditional lecture-based instruction remains common~\citep{henderson2012innovation,dancy2024physics}. 
We also emphasize that the sample reflects a
particular moment in time. AI technologies are evolving rapidly beyond
higher education, and institutional initiatives have likely moved
forward even since the data was collected. As a result, relationships
between institutional context and AI use are likely to change if adoption
becomes more widespread and more formally supported. 

Despite these limitations, the study provides a targeted snapshot of a pedagogically
engaged population likely to be actively experimenting with AI in research and teaching.
The limited presence and effect of institutional initiatives in this dataset reveal 
the signal more cleanly: that significant variations in usage and strong epistemic 
concerns exist even within this population. These patterns may evolve further
as AI technologies and institutional responses mature.

The results and these limitations suggest a number of avenues for
further work. How and whether the instructor's orientation
toward AI drives pedagogical choices and classroom practice is
important to understand, as is whether their orientation transfers to students.
Since the results rely on self-reported data, future
work should examine classroom implementations directly, including
how AI is incorporated into instruction and how students interact
with AI-supported tasks. 

Studies that examine AI in physics learning environments may find
it valuable to take into account our result that different pedagogical
orientations may exist. It would of course be valuable to replicate
this study in the context of STEM faculty more broadly. It is interesting
to consider whether similar stances exist in the student population
as these may shape their responses to introduced tasks. Addressing
the above limitations will require studies that combine observational
methods with fine-grained analysis of student reasoning in AI-supported
tasks.

Finally, the emergence of epistemic considerations as a factor in
adoption, especially AI's potential to enable delegation of disciplinary
thinking, highlights the need for research on how AI usage affects
student thinking. More broadly, these results resurface and underscore
a long-held objective of PER, to examine and facilitate scientific
thinking, or ``doing science,'' in physics classrooms.

\section{Conclusion}
\label{sec:conclusion}

This study identifies a coherent construct, \textit{AI pedagogical
orientation}, that strongly predicts faculty use of AI across research,
teaching, and other professional activities in physics, chemistry
and astronomy. Rather than reflecting simple differences in access,
information, or general attitudes, this construct captures how instructors
interpret the epistemic role of AI, namely whether it is seen as supporting
or undermining disciplinary thinking. Across both quantitative and
qualitative analyses, variation in AI use was most strongly associated
with differences in practical engagement and in how instructors reason
about the relationship between AI and learning.

These findings suggest that existing models of technology adoption
may be incomplete in contexts such as AI, where tools have the potential
to dramatically change how knowledge is produced and evaluated, and
that facilitate delegation of thinking to automated systems. While
factors driving adoption of other technologies such as perceived usefulness,
ease of use, and institutional support remain relevant, and will likely
become more so for later adopters, they do not fully explain variation
in adoption within this population. Existing technology-adoption frameworks help
explain contextual and perceptual influences on AI use, but the present results
suggest that instructors’ interpretations of AI's role in disciplinary thinking
may also play a central role in decisions about whether and how these tools are used.

Although this study focuses on a population of pedagogically engaged STEM faculty who may be more likely than the broader faculty population to experiment with emerging instructional technologies, even in this group there is no consensus regarding the pedagogical value of AI tools. Unlike many evidence-based instructional practices developed through PER and related education research, the educational impacts of generative AI remain uncertain and contested. Nevertheless, the results point to broader questions about the integration of AI in knowledge-intensive domains. In contexts where AI interacts directly with processes of reasoning, inquiry, and knowledge production, adoption may depend not only on the capabilities of the technology, but also on how individuals conceptualize its role in thinking itself. Future work should examine how such orientations develop over time, both in faculty and students, how they are shaped by disciplinary and institutional contexts, and how instructional design can support productive integration of AI while addressing concerns about its impact on learning.

\begin{acknowledgments}
The authors acknowledge funding from the Research Corporation for
Science Advancement under the CottreLLM collaborative. This study was reviewed by the Tufts University Institutional Review Board and determined to be exempt from human subjects review. The authors thanks Drs. Ted Clark, Elaine Short and Nick Seaver for helpful discussion.

\emph{Author contributions:} TJA, TRH, CLV, NS and JJF designed the survey; TJA performed the statistical analysis and directed the qualitative analysis; ID, MW, TRH, CLV, NS and JJF performed coding; all authors participated in drawing conclusions and preparing the manuscript. 

\end{acknowledgments}
\appendix 
\renewcommand{\thetable}{A\arabic{table}}
\setcounter{table}{0}
\section*{Appendix}
\subsection*{Survey questions}
\label{sec:surveyquestions}
\begin{enumerate}
\item {\small What AI tools have you used in your research? }\textit{Chatbots
and Virtual Assistants, e.g. ChatGPT, Claude, Gemini /Coding assistants,
e.g. Copilot, Codeium, Cursor / Writing Tools, e.g. Overleaf AI assist
/ Content Generation Tools / Data Analysis and Chat-with-Data Tools
/ Another tool(s) {[}free response{]} / I haven't used AI tools in
my research}
\item {\small Briefly explain how you are using these tools in your research.
}{\small\textit{Shown only if Q1 indicated use.}}{\small\par}
\item {\small What AI tools have you used in your teaching? }\textit{Same
responses as Q1}
\item {\small Briefly explain how you are using these tools in your teaching.
}{\small\textit{Shown only if Q3 indicated use.}}{\small\par}
\item {\small What AI tools have you used in conducting the rest of your
job, i.e. other than teaching or research? }\textit{Same as Q1.}
\item {\small Briefly explain how you are using these tools in conducting
the rest of your job, i.e. other than teaching or research? }{\small\textit{Shown
only if Q5 indicated use.}}{\small\par}
\item {\small What informed your decision to use AI tools? Did you encounter
any obstacles? }{\small\textit{Shown only if }}{\small\textbf{\textit{any}}}{\small\textit{
of Q1, Q3 or Q5 indicated use.}}{\small\par}
\item {\small What sources of AI tools have you used? }\textit{Free versions
of commercial AI tools, e.g. ChatGPT, Claude, Gemini. / Paid-for or
enterprise versions of commercial AI tools, e.g. ChatGPT, Claude,
Gemini. / AI tools specifically developed for my institution. / Open
source AI tools. / An AI tool I developed myself. / Another tool(s)
{[}free response{]}}
\item {\small Does your institution offer degree programs or certificates
involving AI? }\textit{Yes / No / Not sure {[}for questions 9-12{]}}
\item {\small Has your institution launched strategic initiatives around
AI? }{\small\par}
\item {\small Does your department offer degree programs, courses or certificates
involving AI?}{\small\par}
\item {\small Has your department launched a strategic initiative around
AI?}{\small\par}
\item {\small I would like to use AI tools in my teaching. }\textit{Strongly
disagree / Disagree / Neither agree nor disagree / Agree / Strongly
agree {[}for questions 13-25{]}}
\item {\small My department has encouraged me to integrate AI tools into
my teaching or research.}{\small\par}
\item {\small My institution has encouraged me to integrate AI tools into
my teaching or research.}{\small\par}
\item {\small I am aware of students using AI tools on their own.}{\small\par}
\item {\small Use of AI is primarily for cheating.}{\small\par}
\item {\small I am comfortable using AI tools in my teaching.}{\small\par}
\item {\small I have prior experience teaching computation as part of my
discipline.}{\small\par}
\item {\small Learning to use AI tools is important for undergraduate majors
in my discipline.}{\small\par}
\item {\small AI can help students solve more difficult problems.}{\small\par}
\item {\small AI allows me to introduce more challenging problems than I
could otherwise.}{\small\par}
\item {\small I’m excited by the possibilities afforded by AI.}{\small\par}
\item {\small Integration of AI is attractive to funding agencies.}{\small\par}
\item {\small I’m concerned about integrating AI into education.}{\small\par}
\item {\small What pedagogical techniques do you use in your teaching? }\textit{Computer
simulations or visualizations (e.g. PheT) / Personal response systems,
e.g. Clickers, PollEverywhere, etc. / Peer Instruction / Traditional
lecture / Active learning techniques}\\
\textit{Flipped classroom / I prefer to describe {[}free response{]}}
\item {\small What sources of information do you consult to learn about AI
in higher education? }\textit{Popular news sources / Discipline-specific
news sources, e.g. APS News, C\&E News, etc. / Colleagues in my department
/Institutional support mechanism / Social media, e.g. LinkedIn, Bluesky,
X, etc. / I prefer to describe {[}free response{]}}
\item {\small What is your career stage?}\textit{ Full Professor / Associate
Professor / Assistant Professor / I have another title {[}free response{]}}
\item {\small In what discipline did you get your last degree? }\textit{Physics
/ Chemistry / Astronomy / A different discipline {[}free response{]}}
\item {\small What is the name of your department?}\textit{ Chemistry / Astronomy
/ Physics and Astronomy / Physics / A different department {[}free
response{]}}
\item {\small What is the highest degree awarded by your department? }\textit{Doctoral
/ Masters / Bachelors / Associates}
\item {\small How would you describe your institution? (Check all that apply)}
\textit{A research university / Minority Serving Institution / A primarily
undergraduate institution / Rural / A liberal arts college / Urban
/ Private / State / I prefer to describe {[}free response{]}}
\item {\small If you have any other thoughts around AI in Higher Education
that you feel were not captured by this survey, feel free to enter
them here.}{\small\par}
\end{enumerate}

\subsection*{Initial codes}

A list of initial codes produced in the first stage of qualitative analysis is shown in table \ref{sec:initialcodes}.

\begin{table*}[ht]
\centering
\begin{tabular}{|l|l|}
\hline
\textbf{Code} & \textbf{Brief Description} \\
\hline
Coding & Including any programming task. \\
Visualization & Including figures and explanatory visualizations. \\
AI-development & Active engagement in AI design and algorithms. \\
End-user & Referring to specific AI tools (e.g., AlphaFold, course-specific chatbots). \\
Ideation & Brainstorming or conversations with another scientist. \\
Basic-info & Getting information about unfamiliar fields, similar to an interactive encyclopedia. \\
Resource-gathering & Literature review or identifying relevant software/resources. \\
Writing & Generating text, tone edits, proofreading, outlining, or compressing text. \\
Administrata & Administrative tasks, including spreadsheets and organizational work. \\
Sentiment-positive & Quote expresses a positive or very positive sentiment toward AI. \\
Sentiment-negative & Quote expresses a negative or very negative sentiment toward AI. \\
Generation & Creation of new things, documents, text, or other outputs. \\
Refinement & Improvement or revision of text, documents, or related materials. \\
Empathy & Considering how students might react to content or rewriting difficult emails. \\
Assessing-AI & AI-proofing, teaching students to evaluate AI responses, or effective AI practices. \\
Exploration & Innovation, exploration, curiosity, recommendations, or news discovery. \\
Productivity & Efficiency gains or support provided by AI. \\
Validation & Improving quality or checking correctness. \\
Reliability & Challenges or limitations encountered when using AI. \\
Obligation & Feeling pressure or necessity to use AI to stay informed or competitive. \\
Accessibility & Concerns about cost, access, or limitations of AI systems. \\
Ethical & Ethical concerns or obstacles related to AI use. \\
Learning-concerns & Tradeoffs between reduced independent thinking and increased productivity. \\
Nomenclature & Comments about how AI, LLM, or ML are identified or described. \\
Institutional & Comments about institutions and their role or response to AI. \\
\hline
\end{tabular}
\caption{\label{sec:initialcodes}\textbf{Codebook of AI-related themes and descriptions}}
\label{tab:ai_codebook}
\end{table*}

\bibliographystyle{apsrev4-2}
\bibliography{bibliography}

\end{document}